# Analyzing Social Network Structures in the Iterated Prisoner's Dilemma with Choice and Refusal

Mark D. Smucker[*]

E. Ann Stanley[†]

Dan Ashlock[‡]

University of Wisconsin-Madison, Department of Computer Sciences
Technical Report CS-TR-94-1259

15 December 1994

## Abstract

The Iterated Prisoner's Dilemma with Choice and Refusal (IPD/CR) [46] is an extension of the Iterated Prisoner's Dilemma with evolution that allows players to choose and to refuse their game partners. From individual behaviors, behavioral population structures emerge. In this report, we examine one particular IPD/CR environment and document the social network methods used to identify population behaviors found within this complex adaptive system. In contrast to the standard homogeneous population of nice cooperators, we have also found metastable populations of mixed strategies within this environment. In particular, the social networks of interesting populations and their evolution are examined.

## 1   Introduction

Social interactions are often characterized by the preferential selection of partners. An obvious example is the selection of sexual partners, but there are many others. Researchers carefully choose collaborators. Salmon fishermen in southeast Alaska share information about salmon locations within small, carefully chosen groups, where the payoffs for receiving but not giving true information are high [17]. Male guppies choose partners for predator inspection [10, 11]. Food sharing between families may have been one of the major forces leading to village formation in the American southwest [31].

---

[*]University of Wisconsin-Madison, Department of Computer Sciences, 1210 West Dayton Street, Madison, WI, 53706, e-mail: smucker@cs.wisc.edu

[†]Department of Mathematics, Iowa State University, Ames, IA, 50010, e-mail: stanley@iastate.edu

[‡]Department of Mathematics, Iowa State University, Ames, IA, 50010, e-mail: danwell@iastate.edu

Preferential partner selection occurs partly because it helps minimize risks associated with defection. Sexual partners may pass diseases, lower social status, provide poor genes to offspring, or opt out of child rearing. For salmon fishermen, the cost of sharing with untrustworthy individuals is the loss of valuable fish, and so any individuals who betray that trust are eliminated from the group. Guppies are less willing to interact with guppies who have previously defected by hanging back during predator inspection; a guppy whose partner hangs back is in much greater danger of being eaten. Friends may defect by passing on sensitive information, missing meetings, and many other ways. Orbell and Dawes [41] argue that, since cooperation helps increase total accumulated wealth, and partner selection enhances cooperation, it is to the benefit of society as a whole to evolve social structures which allow individuals to choose their partners.

Partner selection creates social networks of interacting individuals which are pathways for the transmission of diseases, information, and cultural traits. Networks of sexual contacts determine the paths along which a sexually transmitted disease, such as AIDS, can spread, and modelers have shown that different network structures can lead to vastly different AIDS epidemics [32]. For various sorts of cooperative behavior, the detailed sociology of who chooses whom and why is a fundamental question about societies. How do groups form? What bonds them? What role to do key individuals play in the society? Social networks link local social interactions to global societal properties.

Social networks and combinatorial graph theory give a framework for the study of social interactions. Social networks have been an area of research in sociology since the late seventies [25, 33, 9, 47] and have recently come under active investigation by AIDS researchers [32]. Knoke [30] presents a good, brief introduction to social networks. Pollock [42] has discussed social nets in the context of the Iterated Prisoner's Dilemma, limiting his discussion to certain behaviors, and Holland [24] has claimed that networks are an essential ingredient of complex adaptive systems.

In a bottom up approach, we have used the Prisoner's Dilemma as a platform from which to begin studying the connections between local social interactions among individuals, their social network structures, and the global properties which result from these local behaviors. To our knowledge this is the one of the initial applications of the techniques of social networks to complex adaptive systems. In section 1.3 we discuss related Prisoner's Dilemma computer simulation research on population structure.

## 1.1  Prisoner's Dilemma

The Prisoner's Dilemma is a game that allows the study of interaction among selfish agents. In Prisoner's Dilemma, two players can make one of simultaneous two moves – cooperate or defect. If both players cooperate, they each get a payoff of $C$, and if they both defect they each get a payoff of $D$. If one player defects and the other player cooperates, the defector gets the highest payoff, $H$, and the cooperator gets the lowest payoff, $L$. The payoffs obey the relations $L < D < C < H$ and $(H + L)/2 < C$. Figure 1 is the payoff matrix for this paper. In Iterated Prisoner's Dilemma two players repeatedly play each other. The restriction on the payoffs, $(H + L)/2 < C$, prevents two Iterated Prisoner's Dilemma players from obtaining an average payoff greater than cooperation by alternating cooperation and defection. When this condition does not hold, a relationship Angeline [1] termed non-mutual

|          | Player A |          |
|          | C        | D        |
|----------|----------|----------|
| C        | (3,3)    | (0,5)    |
| D        | (5,0)    | (1,1)    |

Player B is on the left side (rows C and D).

Figure 1: Player A receives the payoff on the right side of the comma, and player B gets the payoff of the left side of the comma. $L < D < C < H$ and $(H + L)/2 < C$ and $L = 0$, $D = 1$, $C = 3$, $H = 5$.

cooperation gives higher scores than mutual cooperation. In most of this paper we use a standard Prisoner's Dilemma payoff matrix, but in section 8 we briefly explore non-mutual cooperation for IPD/CR.

The best move to make when playing a one-shot Prisoner's Dilemma is to defect. A defection guarantees a payoff of either $D$ or $H$ of which both are better than the $L$ payoff which is possible if one decides to cooperate. But if both players defect, then they both will do worse than if they both cooperate, and hence the dilemma. Defection makes sense when only playing one game with one opponent. When the Prisoner's Dilemma game is repeated, and the number of iterations is unknown to the players, then cooperation has a chance of arising. This is only true if the players hope to maximize their total summed payoffs and not simply beat a single opponent by obtaining a higher total score, for defection on all moves is still the best strategy if they hope to simply conquer.

In a population undergoing coevolution, an individual wants to obtain a higher average payoff than the other individuals in the population in order to survive and reproduce. An individual which scores higher than its opponents by defecting in each of its pairwise matches is not guaranteed the highest average payoff in the population, for if its opponents are able to cooperate among themselves, and at least some of them defect against it sometimes, they may be able to obtain higher average payoffs. On the other hand, if cooperators are not able to minimize their losses or punish a defector, then the defector can invade a population.

Axelrod pioneered the use of computers for studying the evolution of cooperation and developed a theory of cooperation based upon reciprocity from the round robin computer tournaments he conducted [3]. Axelrod was the first to use a genetic algorithm to evolve complex strategies to play the Iterated Prisoner's Dilemma [4, 5], and soon after that Fujiki and Dickinson implemented a system that evolved LISP coded strategies to play the Iterated Prisoner's Dilemma [16]. Miller used a genetic algorithm to coevolve populations of finite state machines [38] which played round robin tournaments of the Iterated Prisoner's Dilemma with varying amounts of noise, where noise meant that with a certain probability an individual's actions were misinterpreted by his partner/opponent. Miller found that a coevolving population's average fitness would initially take a dip as nasty players eliminated naive cooperators, but with time players capable cooperating among themselves and punishing defectors would evolve and take over the population – evolution of cooperation.

Fogel [13] has also evolved populations of automata that play the Iterated Prisoner's

Dilemma using the technique of evolutionary programming [12, 15] and studied the effects of modifying the payoff matrix while still maintaining the proper relations of the payoffs. Marks [37] discusses and reviews the use of automata in repeated games including the work he has done on Prisoner's Dilemma with genetic algorithms. Further reviews of Prisoner's Dilemma research can be found in [2, 7, 6, 36].

## 1.2 Previous Research

The Iterated Prisoner's Dilemma with Choice and Refusal (IPD/CR) is a model in which agents may choose and refuse interactions based on expected payoffs from the other players and thus allows the study of a more realistic but greatly more complicated system. Choice allows defectors to home in on suckers while refusal allows the same suckers to protect themselves from repeated attack. Whereas in Axelrod's and Miller's experiments, reciprocation of defection is necessary for the evolution of cooperation, in IPD/CR a population of players who always cooperate can use choice and refusal to survive invasion by players who always defect. While previous IPD/CR research [46, 2] has primarily focused on general trends of many runs over different parameter settings, this paper will focus on examining what is happening in specific populations to produce seemingly evolutionarily metastable results.

In the genetic algorithm work of Axelrod [5], often populations arose where the best strategy was not a nice one. The strategy would use an initial defection and then begin cooperation with like players. These strategies act as a sort of password system among the players. Those players which know the password of "initial defect then cooperate" suffer minor losses compared to other players who don't. Players which don't use the password will trigger, in a password using player, strategies which are often all defecting or Tit-for-Tat while the password users end up in repeated mutual cooperation. Robson [44, 45] gives an explanation of the evolution of these defect then cooperate strategies as "handshakes" and Inman [27] talks about "insulars" "using a *behavioral protocol* based on a specific sequence of cooperations and defections." We have also observed similar handshakes in plain Iterated Prisoner's Dilemma experiments and view this operation as an example of a constrained system evolving an approximation to a choice and refusal mechanism. In IPD/CR, such passwords do not seem to come into play as individuals rely upon the choice and refusal mechanisms to obtain cooperation. Choice *and* refusal are important enough to a population's stability that if some mechanism for choice and refusal is not provided, they may spontaneously emerge during evolution.

The large number of iterations in the Iterated Prisoner's Dilemma game relative to the population size and memory capacity of agents in both Axelrod's and Miller's work allows for cooperation to evolve without much trouble. In both Miller's and Axelrod's experiments, players are forced to play an Iterated Prisoner's Dilemma game of either 150 or 151 iterations respectively with every other player in a round robin tournament, and in Miller a player also played itself. In real life people often have the ability to choose and refuse with whom to interact, and because of this ability the time of interaction is not necessarily as great as represented in Axelrod's and Miller's experiments. A fully cooperating IPD/CR population can maintain long-term cooperation while having an average interaction length of only 10.2 iterations. Fogel [14] conducted an experiment which allowed individuals to evolve their length of interaction. A longer mean interaction length for a population was often coupled with a higher average fitness but not always, for the coevolutionary dynamics of the Iterated Prisoner's Dilemma still play a very active role as they do in IPD/CR. The change in length of interaction is only one consequence of adding choice and refusal to the Iterated Prisoner's Dilemma, and other effects have been reported in [2, 46].

The IPD/CR simulations have shown to produce fairly distinct levels of evolutionarily metastable behavior that show up visually in the graphs of the populations' average fitness versus generation, where an individual's fitness is defined to be its average payoff. The different regions are characterized by the average number of defections per individual. While these metastable behaviors are also observed in Iterated Prisoner's Dilemma without choice and refusal, choice and refusal greatly decreases the number of them that are observed, and thus enhances the significance of the ones that are observed. This tendency for IPD/CR populations to cluster in average payoff space is discussed in [2, 46].

## 1.3 Population Structure

Population structures describe the way in which a population of individuals interact with one another. Researchers have tended to look at either spatial or behavioral structures. Both types of structures can either emerge from or be imposed on the population. IPD/CR produces emergent, behavioral structures that we characterize by social networks.

Spatial population structure tends to refer to how real populations are spread out in one, two, or three dimensions. For example, one dimensional populations could exist along a beach front, two dimensional structures often model simple plant populations and land roving animals, and three dimensional structures can be found populations which live in trees or the ocean. Spatial distribution has been hypothesized as a cause of speciation [49].

Behavioral structures describe the behavior between individuals in contrast to simply considering the fact that two individuals separated by a great distance tend not to be able to directly interact. Individuals can be spatially separated in human organizations but be close in a behavioral structure. Take for example an employee who lives in Los Angeles but takes his orders from his boss in New York City. The employee is close to his boss in a behavioral structure that shows the hierarchy of command but is far away spatially.

Many researchers have studied the effects of limiting interactions between social agents by imposing spatial structure on Prisoner's Dilemma populations. Mühlenbein studied the evolution of cooperation in various population structures and demonstrated that cooperation emerged sooner with an implementation based on Darwin's Continent Cycle than in other population structures [39]. Axelrod studied the idea of clustering cooperators together so that they can invade a population of defectors [3]. A commonly implemented spatial structure is a two dimensional cellular automata where each cell is a player, and that player's interactions with the other members of the population is limited to either its four or eight closest neighbors. Nowak and May ran Prisoner's Dilemma on a grid and produced interesting patterns [40], but Huberman and Glance criticized Nowak and May's synchronous updating of the cellular automata – in nature synchronous updating does not correspond to reality – and showed that an asynchronous updating causes the patterns to disappear [26]. But Lindgren and Nordahl [34] investigated a grid based population more as an interesting cellular automata than a real biological or social system and reported that interesting patterns emerge with asynchronous updating of cells if the individuals have greater memory depths. These population structures say who may interact with whom.

In contrast to researcher imposed spatial population structures, behavioral structures tend to be emergent properties of the social system, and it is up to the researcher to measure who interacts with whom. Lindgren and Nordahl examined the formation of food webs within a Prisoner's Dilemma simulation where individuals can use tags to determine who to play. The tags act as a form of attractiveness [35]. Glance and Huberman have done work on fluidity of the population structure and have reported that allowing players to join and leave clusters of others players aids in the formation of cooperation for the N-person Prisoner's Dilemma [18]. While there is some similarity between our work and that of these other two groups, in IPD/CR we are interested in who plays whom and not in who eats whom, and the IPD/CR behavioral structure is not as constrained as that of Glance and Huberman but instead is a measured property describing relationships that emerge from individual behaviors.

Previous research which has addressed the issue of social choice with computer simulations but to our knowledge has not looked at the issue from a structural viewpoint includes Riolo [43], and Batali and Kitcher [8]. Riolo has done preliminary experiments with agents who refuse partners based on players' tags. Batali and Kitcher present a simulation in which agents receive a payoff for opting out of games and while similar to IPD/CR, agents do not choose and refuse social interactions as in IPD/CR, but simply get to decide when to quit playing someone.

Kirman [28] talks about evolving networks of interactions between economic agents. Kirman's idea is distinctly different than our approach here. We allow the network structure to emerge naturally, but Kirman talks about the network structure evolving as a separate entity which is operated on by some algorithm. The evolution of network structures has been done in the artificial neural network field by several researchers [20, 29, 21]. With neural networks, the search for a better topology makes much sense, but if one's goal is to study natural systems then one doesn't want to search for the best connection of agents but instead to study emergent structures.

# 2   IPD/CR

IPD/CR consists of a population of players which are coevolved over some number of generations using a genetic algorithm. Each generation is divided up into iterations. Unlike traditional round robin implementations of Iterated Prisoner's Dilemma in which each player plays every other player each iteration, IPD/CR allows players to choose and refuse player interactions at each iteration.

## 2.1   Choice and Refusal

The players in the IPD/CR simulations studied here consist of a sixteen state Moore machine which is coded as a one hundred forty-eight bit string for use by the genetic algorithm as in [38]. In Ashlock et. al [2] and Stanley et. al [46] we also used a Mealy machine formulation and have found that the behavior of IPD/CR is not terribly sensitive to this representation issue. The Moore machine is used to play the Iterated Prisoner's Dilemma and is the only part of the player structure which evolves. A player $n$ maintains an expected payoff $\pi(m|n)$ of every other player $m$. $\pi(m|n)$ is used to determine who is best to play in the current iteration and who is intolerable. Another player $m$ is intolerable if $\pi(m|n) < \tau$ where $\tau$ is the minimum tolerance level. Expected payoffs for the entire population are initialized to $\pi_0$.

Every iteration, each player makes at most $K$ offers of game play to the players from whom it expects to obtain the highest payoffs. In other words, a player chooses the players which correspond to the top $K$ expected payoffs. If more than one player has the same expected payoff, random draws break the ties until $K$ offers are made. If there are less than $K$ tolerable players to choose from then all tolerable players are chosen. If all other players are considered intolerable by a player, then that player receives a wallflower payoff $W$.

For each chosen opponent, a Prisoner's Dilemma game is played between the player and the opponent if the opponent does not refuse the offer of play. If the opponent had also chosen the chooser, only one game is played between the pair and not two. If a player's offer of play is rejected, that player receives a rejection payoff $R$ from the rejector, and the rejector is not penalized. The rejector does not receive a payoff from the chooser. If a Prisoner's Dilemma game was played, each of the two players receives either a $L$, $D$, $C$, or $H$ payoff as determined by the rules of the Prisoner's Dilemma game in section 1.1, and each player changes its Moore machine state accordingly. A player stores a unique Moore machine state for every opponent.

When a player $n$ receives a payoff from an opponent $m$ either by playing a Prisoner's Dilemma game or by being rejected, $n$ updates its expected payoff of $m$ via the following assignment rule,

$$\pi(m|n) \leftarrow \omega\pi(m|n) + (1-\omega)U \quad ,$$

where $U$ denotes the payoff $n$ received from $m$ and $\omega$ is the "memory" weight.

## 2.2 Evolution

A genetic algorithm [23, 19, 48] is used to coevolve the population of players. For this experiment, each generation consists of one IPD/CR tournament. We used a population size of thirty. At the moment, IPD/CR requires that each player remember what state it is in with respect to all other players, and as it is common for humans to remember their current feelings towards large numbers of people, this does not seem like a bad assumption. A generation/tournament is over when the predetermined number of iterations has elapsed. A player's fitness is determined to be the average payoff per payoff it received. At the end of a generation, the top twenty individuals as ranked by fitness, the elite, are chosen to survive to the next generation. Individuals of equal fitness have equal probability of surviving. From the twenty elite individuals, ten individuals are chosen with replacement via fitness biased selection to pair up, mate and fill up the ten openings with their children. An individual can mate with itself. When two individuals mate, their bit strings are subjected to one point crossover and the resulting two children are then subjected to mutation. The probability of a single bit mutation for these experiments was set at 0.005. All individuals' memories of previous plays are removed, so that their expected payoffs are reinitialized to $\pi_0 = 3$ for the next generation, and their Moore machines are reinitialized to their starting states. The number of iterations per generation was set at one hundred fifty.

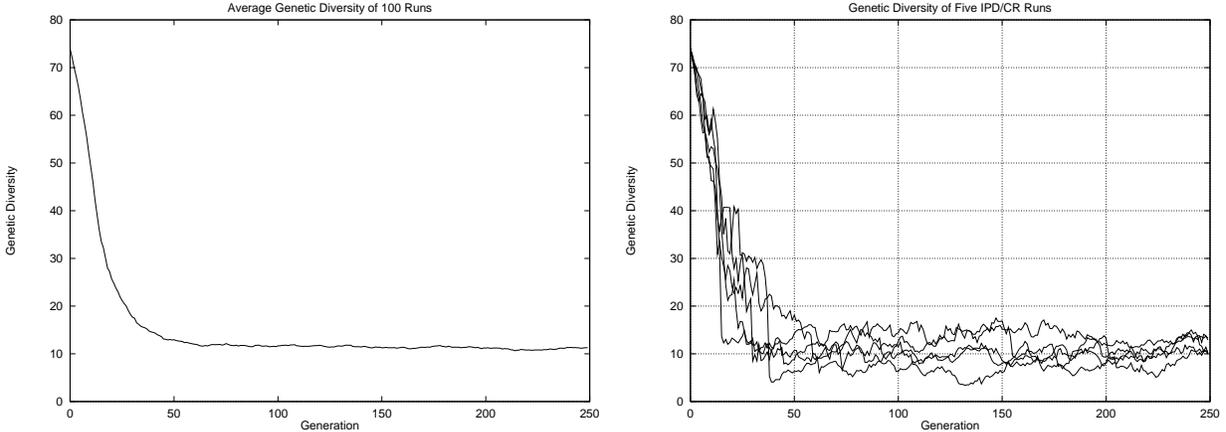

Figure 2: The left figure shows the average decrease in genetic diversity of the 100 runs in this paper. The figure on the right shows five example runs and their genetic diversity over 250 generations. Genetic diversity is defined to be the average of players' average hamming distance to other players' bit strings.

| Type of crossover | One point | $\pi_0$ (Initial expected payoff) | 3.0 |
|---|---|---|---|
| Probability of crossover | 1.0 | $\tau$ (Minimum tolerance level) | 1.6 |
| Probability of a bit mutation | 0.005 | $K$ (Maximum number of offers of play) | 1 |
| Number of iterations | 150 | $R$ (Rejection payoff) | 1.0 |
| Population size | 30 | $W$ (Wallflower payoff) | 1.6 |
| Number of elite | 20 | $\omega$ (memory weight) | 0.7 |

Table 1: The parameters for the populations examined in this report.

Initial experiments have shown that while the specifications of the genetic algorithm affect the outcome of the experiments, the social network results of the experiments are general over many parameter settings. When a successful mixture of players evolves, the genetic algorithm quickly creates a population which is fairly homogeneous genetically. Figure 2 shows the decrease in genetic diversity for the runs in this paper. However, because of the way the Moore machines are coded as bit strings, a single bit mutation can create an individual which behaves drastically different from its parents, and thus the genetic algorithm can create "invaders" of the semi-stable elite population. The dynamics of IPD/CR cause the populations to maintain low average ages, two and a half to five generations old. Low average age of individuals implies that there is a constantly changing population, and this allows for drift in unexpressed behavior.

Table 1 describes the IPD/CR environment examined in this paper. A study of other IPD/CR parameter settings is reported in [2].

# 3 Understanding Population Behaviors

Despite the simple formulation of IPD/CR, the population behavior which results is very complex, and varies significantly between runs with different random seeds. This complexity is a desirable feature of the model, as it is with any artificial life system which seeks to capture some of the richness and variability of the real world. In order to understand such a system, we need to examine it from a variety of perspectives. This is a new field, and tools for doing this research must be borrowed from other fields, or developed from scratch. We have developed and appropriated a number of such tools. The results of the studies presented here show that we need even better tools: we believe that most multiagent studies will benefit from some of the approaches we demonstrate.

To begin with, we can look at IPD/CR from a global perspective using the population statistics of average payoff, minimum payoff, maximum payoff, Hamming distance between player's encoded bit strings, etc. Such a perspective can give general information but almost always cannot say anything about why the population has such statistics. At the other extreme, we can look at each individual's Moore machine. This local perspective tells us how an individual is constructed and how it could interact with any other individual. However, neither the global nor this myopic local perspective gives enough information for us to understand our system. An individual's expressed behavior depends upon the behavior of the other individuals present in the population, and the global characteristics like average payoff mean little without understanding the local social interactions.

In IPD/CR, individuals choose their own interactions based on their opinion of other individual players. These interactions build up networks of interacting players, and the results of these interactions lead to the global statistics. Thus one of the most crucial aspects of our simulations is the network structure, which is intermediate between the global and the local. Techniques used in the theory of social networks help to bridge the gap between global and local by using combinatorial graph theory to characterize the social interactions in the population and provide a view of the whole.

We next describe our social network method, the significant play graph, and then the visualization system used to investigate IPD/CR populations.

## 3.1 Significant Play Graph

A network or combinatorial graph [22], from now on referred to simply as a graph, is a useful abstraction of social interactions. A graph consists of a set of *vertices* or *nodes* and a set of *edges*. An edge connects two vertices. In terms of social networks, the vertices of the graph represent individuals, and the edges represent some property such as friendship that the two individuals share. Edges can also be assigned a weight. The edge weight can be used to represent the strength of the property between two nodes.

Traditionally, social network research has dealt with directed graphs, *digraphs*, in which a direction is placed on the edges. Digraphs would, for example, allow the study of flows, and in a social sense, digraphs could show us that John likes Mary, but Mary doesn't like John. However, digraphs add an extra layer of complexity, and we could not make use of this information at this point in our studies. Therefore, for simplicity, we will use undirected graphs in this paper, in which vertices are individuals, and edges represent significant play between two individuals.

The construction of the IPD/CR significant play graph consists of several steps. First, at the beginning of each generation, all of the edges of a complete graph (each node is connected by an edge to every other node except itself) are given a weight of zero. When two individuals play a game of Prisoner's Dilemma, the weight of the edge connecting the two players' nodes is incremented by one. The maximum weight an edge can have is equal to the number of iterations per generation, one hundred fifty.

At the end of a generation, the distribution of edge weights can be plotted in a histogram to show general population characteristics. A population in which the individuals randomly play each other will have a bell-shaped distribution of edge weights with a mean of around ten and a standard deviation of 3–4, but a population that contains members which have strong preferences for a few specific members of the population will show a distribution at the higher end of the spectrum with a second high peak of near zero weighted edges between players that rarely interacted. While the distribution can show general population characteristics, it does not show who played whom. To get a better understanding of what is going on in a population, the edges of the play graph which do not represent significant levels of game activity are zeroed out and the other edges have their weights set equal to one.

For purposes of the current analysis, an edge is considered significant if it is greater than a threshold value. If the standard deviation, $\sigma$, of the edge weights is greater than the mean edge weight, $\mu$, then the threshold is set equal to $\mu$, otherwise the cutoff is set equal to $\mu - 2\sigma$, unless this causes the threshold to be negative in which case it is instead set equal to one. After applying the computed threshold to the edge weights, the edges with a weight of zero are considered to be deleted, and the remaining edges determine the significant play graph.

The determination of what represents significant play is the key to obtaining useful graphs. What appears to be a significant connection depends in part on the population being observed. We have tried a number of procedures, visually comparing the results to the observed behavior of the individuals (see below in the visualization section). An obvious choice is $\mu - 2\sigma$, but in populations that have great variation in edge weights, this choice results in a negative threshold and a fully connected graph. On the other hand, setting the threshold to $\mu$, while eliminating this problem, cuts off too many connections in other populations. Fixed thresholds, while they give nice results for specialized population types, are not sufficiently general. The current, somewhat ad hoc, procedure was chosen so that single contacts between individuals are usually not counted, and secondly when a significant portion of the edges are distributed together, forming a single mode, noisy and low weight edges are set to zero, and thirdly when there is a bimodal distribution where the edges in the lower mode are at reasonable levels some will be kept, but if they are very low they will be cut off. This gives results which capture much of the structure seen in dynamic visual observations, but more work needs to be done on this issue. Future plans include the exploration of methods which favor interactions which occur later in the generation over those which occur earlier, since we are interested in the long-term social network structures which arise in populations. Interactions early in the generation are dictated by the agent's attempts to learn about their environment and rarely reflect long-term behavior.

The significant graph measures of the *average degree*, *maximum degree*, and number of *connected components* can give insight to a population's behavior. The average degree and maximum degree are calculated from the degrees of all of the nodes, where the degree of a node is equal to the number of edges to which it is connected. A graph is connected if all vertices are reachable by all other vertices via a path along the edges which connect the vertices. If a graph is not connected, then it is composed of two or more connected components, where each connected component, often referred to simply as a component, has no edge connecting any of its vertices to any other connected component. The average degree represents the average number of significant relationships per individual. The maximum degree can be used in combination with the average degree to discover populations with central players who are more popular than other players. The number of components can be used to identify populations in which players have clustered into separate groups, or populations that have ostracized individuals, since an ostracized individual often forms its own component.

## 3.2   Visualization

In a complex adaptive system like IPD/CR, visualization of simulations is crucial. Population and significant play graph statistics are sometimes misleading and difficult to interpret. To validate and understand these results, we constructed a simple, interactive, real-time animation.

In this paper we display two types of snapshots from the visualization system. The first is a snapshot of the animation which is an overlap of five iterations of play and will be discussed in detail below. The second is the final display of the significant play graph for a generation.

The visualization works as follows. At the beginning of each generation, the twenty elite individuals are placed, starting at "three o'clock," counterclockwise around a circle with their rank in the previous generation determining their order. Completing the circle, the ten children are then placed randomly into the remaining ten spots. During the animation, an individual is represented by an open circle, an individual receiving a rejection payoff is represented by a square, and a wallflower player is represented by a solid circle. Every individual is connected to every other individual via an abstract damped spring. The dynamics of the springs are affected by how often individuals play each other and the number of iterations elapsed. The spring strength becomes stronger between players which play each other. When individuals get too close to each other, they are repelled via a force inversely proportional to their separating distance. The overall effect of this system is to pull players into clusters or bring out the features of central individuals by placing them in the center of a group. Rejected players, now marked with a square, move away from the rejector. When two players interact, because one of them makes an offer to the other, a line is drawn between them. When a player becomes a wallflower, i.e. it finds all players intolerable, it turns into a solid circle.

To ease visualization, we overlap five iterations of a generation at a time onto the display. The animation can be paused at any point, and then the researcher can use the mouse to point and click on agents to get statistics on them and display their Moore machine structure. One can also click on a pair of individuals and watch their pair play and expected payoffs change over twenty iterations or until one finds the other intolerable. We have found that viewing the behavior an individual expresses with others is the most useful of these two tools, as examining sixteen state Moore machines is tiresome and often gives little insight into behavior within an individual's own population.

At the end of each generation, the individuals in the system are arranged back on their starting circle, the individuals' fitnesses are listed next to them, and the significant play graph is drawn. Note that at this point wallflowers and rejectees are not given any special designation.

# 4    Individual Behavior

One of the fundamental dogmas of artificial life is the idea that local interactions among autonomous objects can lead to global organization. As we detail later, organized population structures do form and persist in IPD/CR, but in order to understand these organized structures, we first we need to look at the local interactions among IPD/CR agents which create the population structures.

Individual behaviors in IPD/CR are most obviously determined by the choice and refusal parameters, their Moore machines, and the structure of the choice and refusal procedure. However, these things determine only the potential behavior of the individual in any population. A player's actual behavior in a population is a product of the environment in which it is placed. A group of players may all cooperate with each other, but if a mutant is placed into their population, each one of them may react to that mutant in completely different ways. The interconnections of IPD/CR are such that this one mutant could potentially change other players' "feelings" about one another, and thus who ends up playing whom. This would then change the social network structure of the population, and thus its global properties.

Players' "feelings" to one another can be viewed as "liking" or "not minding" or "disliking" another player. A player likes another player that has the highest expected payoff. Liking a player will cause the attracted player to choose the player it likes. A player does not mind another player as long as that player has an expected payoff greater than or equal to the minimum tolerance level $\tau$. Not minding another player means that a player will accept game offers from players it does not mind. A player dislikes another player if that player's expected payoff is less than $\tau$. If one dislikes another player, one will refuse game offers from that player.

From these three "feelings," players form organized persistent social structures. The sensitivity to their "feelings" which allows mutants to affect their behavior occurs because the choice mechanism is finicky in the sense that if the expected payoff from two different players is very small, a player will always choose the player with the higher expected payoff. The players have no "fuzziness" in their selection. Hence, players may get stuck in the social equivalent of a local optima.

In the following two sections we will describe a number of organized population structures we have observed which can be described by the individual behaviors of which they are composed (see Table 2).

| Full Nice Cooperation | Everyone likes one another. |
|---|---|
| Latching | Individuals like only a few other individuals, but don't mind the others. |
| Raquel and the Bobs | Bobs like all Raquels. Raquels like each other and don't mind Bobs. |
| Stars | Hubs dislike one another. Spokes like hubs. Hubs don't mind spokes and like the spokes they are connected to in sequence. |
| Connected Centers | Nice guys like each other and don't mind the thugs. Thugs like and latch onto a center nice guy. |
| Wallflower | Everyone dislikes one another. |

Table 2: A summary of the population behaviors described in sections 5 and 6 in terms of the individual behaviors which act to give the populations their characteristic social networks. The terminology is defined in the relevant subsections of sections 5 and 6.

# 5    Population Behaviors

A large number of population structures occur in IPD/CR. Generally no single behavior emerges and persists forever, indicating that populations which are stable against mutation and crossover probably don't exist. However, there do appear to be general population structures which can be thought of as metastable in the sense that they emerge frequently and persist for many generations. We will define and describe five of these population structures in detail below: full nice cooperation, latching, Raquel and the Bobs (R&B), stars, and connected centers. These populations, each of which we have repeatedly observed for the environment discussed in this paper, distinguish themselves as organized and often persistent over many generations.

The IPD/CR populations often make rapid transitions from one of these metastable behaviors to another, sometimes after spending a number of generations with a very low average fitness. Characterizing all of the ways that this could happen is infeasible, but in section 6 we will describe one transitory situation, where the population average fitness *crashes* (drops rapidly to a low value), and the population spends a small number of generations in a wallflower ecology before moving into full nice cooperation.

The observations we present are drawn from one hundred runs of two hundred and fifty generations each of the current IPD/CR simulation system described in this paper, and several hundred runs of previous implementations both with and without choice and refusal.

## 5.1    Full Nice Cooperation

A population exhibits full nice cooperation when each individual cooperates in every Prisoner's Dilemma game. A player is nice if it cooperates on its first move. Since all players maintain an expected payoff of three from all other players, partners are always chosen at random. The play graph of a population characterized by random choice is simply a fully connected graph that has average and maximum degree equal to twenty-nine and a single connected component.

At the parameter settings of the current paper, approximately three quarters of the simulations end up in full nice cooperation, which tends to persist once it is established. Usually when invaders appear, the population successfully ostracizes them and maintains a near three average payoff. However, our investigations indicate that no population will stay in full nice cooperation forever. If followed for a large enough number of generations, every fully cooperative population will eventually be invaded by less cooperative individuals. In order for a fully cooperating population to withstand invasion, it needs to very quickly refuse play from nasty opponents and stick them with low rejection and wallflower payoffs. Random mutations of the portion of the genome unused by a fully cooperative population ensures that eventually some portion of the population will lose its ability to protect itself against defectors. Even defensive structures used during a past incursion of uncooperative players may be lost if these structures are no longer being used.

Later in section 5.3 we describe a population behavior which causes a peaking in the population's average payoff, but we also see at times a situation where the cooperating population's payoff dips. These dips last longer than the frequent small drops in average payoff caused by random mutants and is caused by a type of player which delays its defection several moves into cooperating play with other individuals and can avoid receiving very many rejection payoffs before the end of the tournament. Because a full nice cooperation population typically only plays on average 10.2 games between every player, it is possible for a player to count the approximate end of play between individuals with its Moore machine and defect on the last few moves. However, as this new player reproduces in future generations, its success is its demise as there are no longer enough naive cooperators in the population, the new "delayed defection" players become extinct, and the population returns to full nice cooperation.

## 5.2   Latching

The choice and refusal mechanism allows individuals to latch onto other players. A latcher is a player who repeatedly chooses to play one or possibly a small number of other individuals. Latching occurs because the expected payoff to the latcher from the latchee becomes greater than it is from the rest of the population and then it stays greater for the majority of iterations. The play graph of a population of latchers usually has three to twelve components, an average degree between one and two, and a maximum degree between two and ten. The distribution of edge weights is such that there are few edges, but existing edges are heavily weighted with weights around 130–145. An example significant play graph and five iteration snapshot for a population of this type can be seen in Figure 3.

The most common form of a latching population is one in which players initially defect for one or two moves and then begin cooperating with each other. Take for example a population of players that have an initial defection followed by cooperation. The Bobs (section 5.3) are a typical example of such a population. Before any play, the individuals all hold an expected payoff of $\pi_0 = 3$ for all other players. Each individual will then choose one player at random from among this group and a game will be played between those two players. Since both will defect, each player's expected payoff of the other play partner will drop from 3.0 to 2.4. All individuals which have not played each other yet will still have an expected payoff of 3.0 from each other. All players will continue to choose from the set of players they have not yet played until some individuals have played everyone who will now have an expected payoff of 2.4 from everyone, and they will make their next player choice at random, both players will cooperate with each other, and their expected payoffs from each other will rise above 2.4. At this point any individual who has already played everyone prefers those it has played twice to those it has played once, and it will select one of these partners on the next iteration. Since its estimation of a partner will now continue to grow the more times it plays it, it will latch onto the player or small set of players it has played the most times.

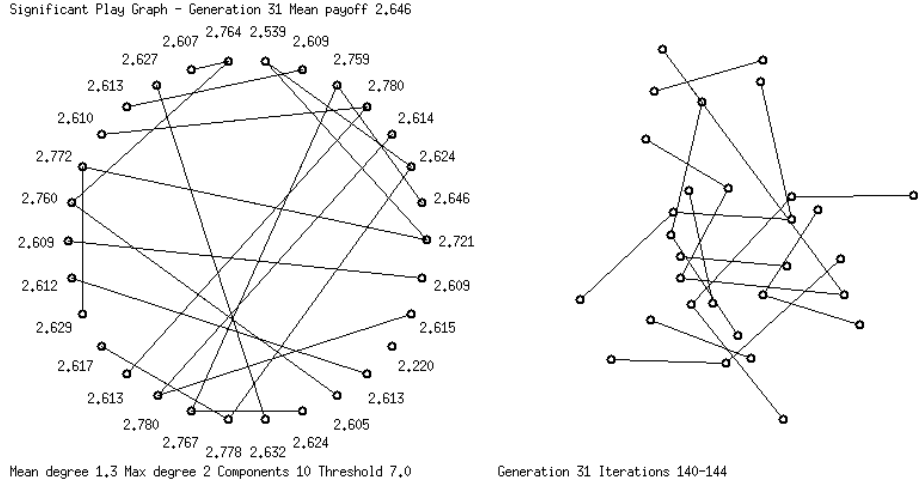

Figure 3: A population of latchers' significant play graph for an example generation and a snapshot from late in the generation. These players have an initial move of defect and then cooperate. This is from the same run as Figure 5.

Interestingly, the random selection process at the beginning, together with the fact that those who have not yet played everyone will prefer those they have never played to those they have ever played, implies that all players will not reach the point of being ready to latch onto a fellow player at the same iteration. Because of this, players who finish earlier will latch onto other members of the population and once their latchees have played all players once they will prefer their latcher(s) to all others rather than choosing their next partner at random. This leads to a very different network structure than if everyone were to select a partner at random and then stick with that partner choice. 100 runs of a pure latching population of 30 individuals and a population which latches onto the first partner selected gave the results in Figure 4. These results look different than the observed number of 3-12 components in the IPD/CR simulations because those populations often contain some mutants who are not pure latchers.

Variations on the above behavior occasionally occur. Table 3 shows a situation we observed once in which two players play for four iterations by first playing two defections each then two cooperations each and on the fifth move player A gets a $H$ payoff from the other player and then three iterations later trades attraction back to player B. These players continue this behavior for all 150 iterations. While the direction of attraction switches between the two players, they have the same social network structure as the more typical defect once or twice then cooperate latchers.

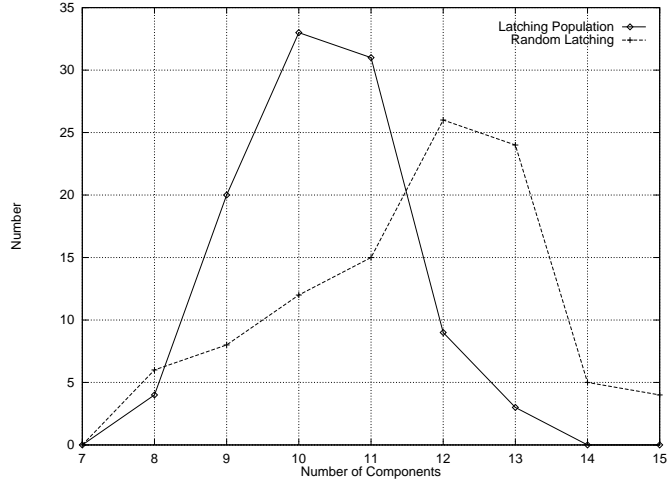

Figure 4: The difference between a latching population and one with random selection of a lifetime partner choice. The mean and standard deviation of the latching population were 10.3 and 1.10, and 11.6 and 1.72 for the random latching. One hundred simulations were conducted of each.

| Player A | ddcc**d**ccccc**d**ccc... |
|----------|---------------------------|
| Player B | ddccccc**d**ccccc**d**... |

Table 3: Two players which exhibit a latching behavior by alternating their attractions. The bold defections show when one player becomes attracted to the other player.

## 5.3 Raquel and the Bobs

Many populations show a peaking behavior in the population average payoff at some point in their evolution. An example of this peaking behavior is shown in Figure 5. The population average payoff increases over a series of generations until it gets very close to 3.0, and then drops to 2.7, or lower, in a single generation. When this behavior is observed in the population average payoff, the population structure Ashlock et al. [2] called Raquel and the Bobs (R&B) is often occuring. Prior to the increase, there are one or more generations of a typical latching population. At the start of the increasing phase of the population average payoff, the maximum degree jumps to 27–29, the average degree stays near 1 and the number of connected components drops to 1–2. The average degree then slowly increases, while the maximum degree and number of connected components are roughly constant. When the drop in average payoff occurs the population's statistics return to that of a latching population. This pattern may occur only once, or it may repeat several times in a row as in Figure 5.

To understand the reasons underlying this behavior, we need to look more closely at the players and their interactions. R&B is best described as the intertwining of two interaction patterns that arise from one another repeatedly. This pattern can occur with different individual behaviors, but the general sequence of events is the same.

The Bobs are often a latching population which play an initial defect followed by cooperation with each other, but we do see varying types of Bobs populations. For example in Figure 5, the first Bobs-like population appears at generation 18, and consists of at least two subgroups with different genetic material, one of which is more defecting than the other. Raquel is a player that always cooperates when it plays another Raquel and is more co-operative in Iterated Prisoner's Dilemma play against a Bob than the Bobs are with each other.

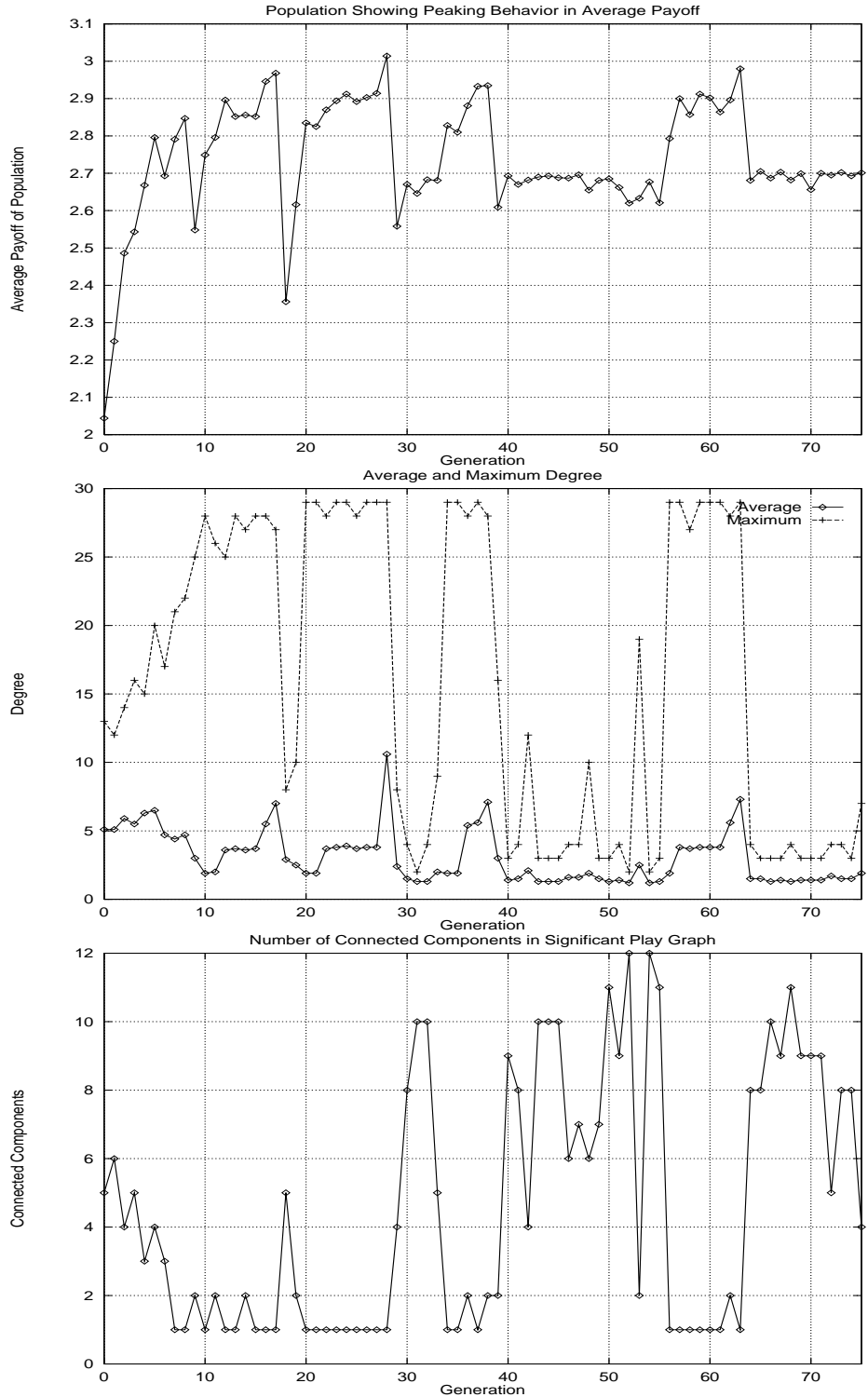

Figure 5: The top figure is an example run showing peaking behavior in the population's average payoff. The middle figure is the maximum and average degree of the significant play graph. The bottom figure is the number of connected components in the significant play graph. Generations 18–75 in this figure show the repeating R&B phenomenon, the first 18 generations exhibit a common settling down-behavior as the population move away from the initially randomly-chosen genetic material.

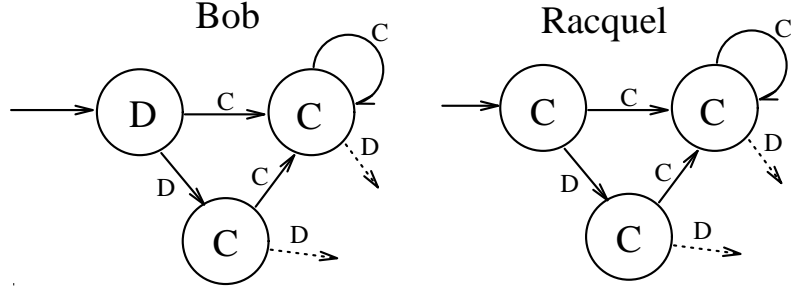

Figure 6: An observed example of Raquel and the Bobs. The only states of Bob's and Raquel's automata that are used when they play each other. Note how Raquel differs only in its start state.

Bobs populations have the extra property that they can produce a Raquel by either a single point mutation or crossover which not all latching populations can do, and so the genetic structure of the Bobs present in the population becomes an important factor. Figure 6 shows an observed example of how a Raquel can be formed by the point mutation of the output of the starting state of a Bob. Understanding the formation of a Raquel from the crossover of two Bobs requires knowledge of the genetic representation of the Moore machines. A player's bit string begins with four bits that code for the starting state of the Moore machine. The remaining bits are composed of sixteen regions of nine bits each that code for each state in the machine (zero through fifteen). One observed example of Raquels via crossover came from a population that contained two types of Bobs which differed not in their expressed behaviors with each other or Raquel but in their genetics. Type A had its start state bits pointing to state nine and type B used state two as the start state. Interestingly, the type B's states nine through fifteen coded for a Raquel if state nine was considered the start state. Thus out of the one hundred forty-eight total bits, there existed a contiguous, eighty-one bit region in which crossover could occur between these two types of Bobs and allow a Raquel to be born.

Since whenever a Bob first plays Raquel it attains a high successful defection payoff of 5, the Bob's expected payoff from Raquel goes from 3 to 3.6, and Raquel's expected payoff from the Bob drops from 3 to 2.1. Raquel has made the Bob fall in "love" with it via its altruistic move and the Bob latches. Driven by its disappointment at the low payoff score it achieves from each initial encounter, Raquel in turn cycles through the Bobs seeking a suitable Prisoner's Dilemma game partner. In this way all Bobs eventually encounter and latch onto Raquel. Note that since a Bob only defects against Raquel once, Raquel will never reject the "obsessed" Bob. After the initial defection against Raquel, the Bobs cooperate with Raquel.

Despite the initial low payoff received from each Bob, Raquel recoups its losses by receiving a large number of cooperation payoffs, getting an average of nearly 3. The Bobs actually fare worse on average, for during the time it takes them to discover Raquel, they are defect-ing with one another. Raquel gets $n-1$ defections, one from each Bob, and then gets one cooperation per iteration per Bob on every iteration after the Bob's initial defection. Since Bobs play roughly two games each round (the one they choose and the one that chooses them) it takes roughly 7 games for each Bob to find Raquel in our 30 player simulations. With 150 iterations this means the fitness of a Bob is about $\frac{15 \cdot 1 + 143 \cdot 3}{157} \simeq 2.83$ while Raquel's fitness is $\frac{29 \cdot 0 + 142 \cdot 29 \cdot 3}{29 + 142 \cdot 29} \simeq 2.99$. Raquel therefore usually receives the highest average payoff in this population and thus one of the highest probabilities of being selected as a parent.

Within a few generations there are multiple Raquels. In Figure 5, the first Raquel is born in generation 20, the second Raquel is born in generation 22, and by generation 28 there are 6 Raquels. The changing social networks for these populations are illustrated in Figure 7. When multiple Raquels exist in the population, the Bobs quickly find all of them and then cycle through them, while the Raquels choose each other. Now the number of times each Bob plays each Raquel is fewer than with only one Raquel in the population. This lowers the average payoff score attained by each Raquel. Conversely, each Bob now spends less time finding the Raquels, which increases its average payoff score.

Herein lie the seeds of Raquel's destruction. As the number of Raquels in the population increases, the average payoff score of the *population* tends to rise, but the average payoff score of a *Raquel* keeps falling relative to the average payoff score of a Bob. Eventually a point is reached where the Raquels score lower than the Bobs (and the Bobs score fabulously well, sometimes, as in generation 28, enough to give a population average greater than 3). This point usually occurs when there are less than 10 Raquels and elitism ensures that all Raquels die out in the genetic step, although sometimes one or two Raquels remain and the growth and decay reoccurs immediately. Otherwise the next population consists only of Bobs.

Figure 7 shows the birth to death cycle from the social net perspective. A population with multiple Bobs and a single Raquel forms a "star" social network with each Bob connected only to Raquel at the center. This network is a singly connected component and has a maximum degree of 29 and a low average degree. The significant play graph shows that Raquel indeed outscored all of the Bobs. As multiple Raquels are born, each Bob is connected to all the Raquels, and the Raquels are connected to each other. The Raquels continue to outscore many of the Bobs, but the difference decreases as the number of Raquels increases. The number of connected components stays one, unless a mutant appears which is unable to play well with anyone and is ostracized. The maximum degree stays 28–29, but the average degree goes up until the population suddenly collapses and loses all its Raquels.

In many of our runs the R&B cycle reoccurs one or many times. The average payoff scores achieved by this succession of populations exhibit persistent spikes as the Raquels emerge, increase, decline to oblivion, and emerge once again. In the case of Figure 5 Raquels reappear again in generations 34 and 56, but in this particular run, after generation 63 the Raquels never again appeared and by generation 80 this population was in full nice cooperation. We have also observed cases where R&B reoccurs many more times than this.

Another interesting phenomenon is that by generation 40 the Bob population is behaviorally uniform, with all Bobs playing a defect followed by cooperation with one another. We conjecture that the spikes in scores and the sharp shifts in population behavior act to eliminate individuals who behave differently from the most successful individuals, and hence help stabilize the population.

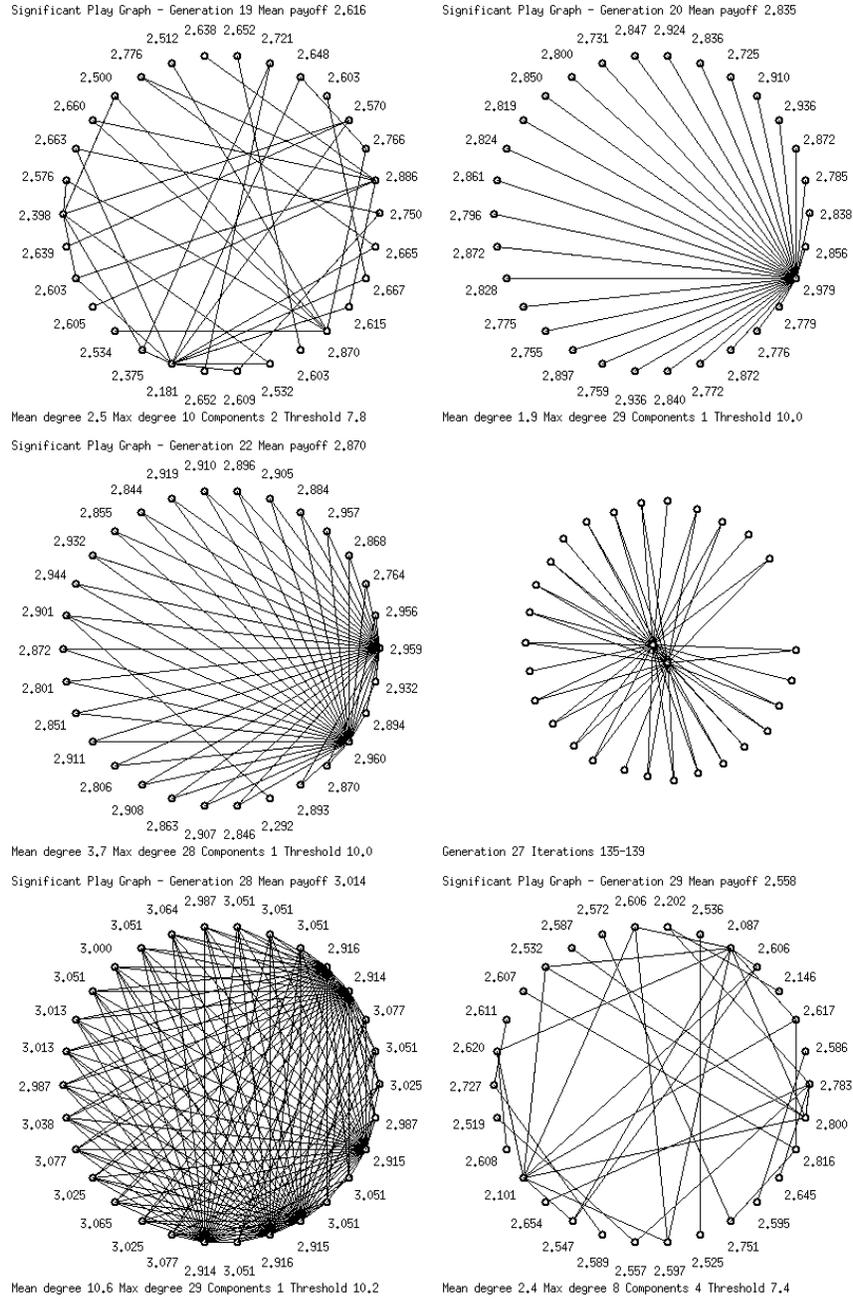

Figure 7: The social networks for the Raquel and the Bobs of Figure 5. From left to right and top to bottom: (1) The latching network of the Bobs population in generation 19 just prior to the birth of the first Raquel. (2) The first Raquel is born in generation 20 and all of the Bobs choose it. At 2.979, Raquel has the highest fitness. (3) Two generations later a second Raquel appears, plus one misfit. The Bobs like each of the Raquels equally well and the Raquels choose each other. Both Raquels outscore all the Bobs. (4) A snapshot from generation 27 with two Raquels, showing how the interactive visualization places them at the center of the net. (5) In generation 28 there are 6 Raquels, and the fitness of the Bobs rises above theirs (and even above 3). (6) The Raquels have died out and the population has returned to latching.

## 5.4 Codependent Populations

In codependent populations two genetically different subpopulations arise and interact with each other. The subpopulations are codependent because the fitness of one population increases as the size of the other population increases. This increase in fitness leads to an increase in that subpopulations' size and thus to a decrease in its fitness. Thus the number of individuals in each subpopulation oscillates back and forth somewhat like a predator-prey cycle. We have coined the term codependent because this case is more benign than predator-prey since *both* subpopulations do better when there are more of the other. Often codependent populations are extremely stable against invasion from their own offspring, perhaps as stable as cooperating populations, and they sometimes persist until the end of the simulation. We see two distinct types of codependent populations: one which looks like disconnected stars and one which looks like a central connected group surrounded by individuals each connected to one of the central individuals.

### 5.4.1 Disconnected Stars

Disconnected stars, or stars for short, are caused by the interactions of two subpopulations, hubs and spokes, which stay in equilibrium with each other for many generations. Figures 8 and 9 show this pattern for an example run in which stars appeared. The significant play graph of a star population consists of disconnected groups, each of which has a single center individual (a hub) to which all of the other individuals in the group are connected, and several (usually 3–7 ) outside individuals (the spokes) who are not connected to each other. The play graph thus looks like disconnected stars, or hubs with spokes. The number of hubs and spokes tends to oscillate up and down somewhat, and the exact balance depends upon the play behaviors of the two. Roughly speaking, star populations are characterized by a moderate average payoff which stays between 2.4 and 2.7 as the balance between the two populations shifts, 1–8 connected components, and a low average (1–3) and moderate maximum degree (5–19). These global measures are similar in behavior to that of a latching population, although stars tend to have slightly fewer connected components and a slightly greater gap between the maximum and average degree. As well, if only a single hub happens to be present for several generations, which may, for example, occur if the original population was a latching population and a hub was born, then the play graph will temporarily look the same as that of a R&B population. These similarities make an algorithm for automatic population classification difficult to develop based on only these measures.

There are many individual behaviors which lead to the formation of stars. What is necessary is that the hubs, while being nice at first, eventually retaliate against the spokes. As well, the hubs are not nice to each other. Table 4 shows the interactions of the individuals in the population of Figures 8–9 with each other.

Given the parameters of the problem, the spokes then always prefer a hub which they have played once (and for whom the minimum expected payoff is 2.52) to another spoke which they have also only played once (expected payoff 2.4). The hubs will end up preferring spokes to other hubs because of the one high payoff they get from them. The hits that the hubs take, and their choice behavior, together ensure that the spokes end up preferring one hub over the other and do not alternate between them.

Star behavior is often not as clearcut as R&B. Sometimes some of the spokes are playing each other as well as the hubs. This decreases the number of connected components. As well, often the hubs are not all identical.

### 5.4.2 Connected Centers

This case is similar to that of disconnected stars, except that the centers, who we shall refer to as the nice guys, always cooperate with each other, and thus like each other. This creates a significant play graph with a completely connected central subpopulation and a second subpopulation, referred to as the thugs, whose individuals are each connected only to a single node in the central group. Except when an occasional mutant is present, the significant play graph is a single connected component, with a maximum degree oscillating around 10–15 and an average degree of 3–5.

These populations have the interesting property that the nice guys would be better off without the presence of the thugs, who are taking advantage of them, and yet they cannot do without them once they are present. An example of the interactions between the individuals in this population with each other are described by Table 5. The nice guys cooperate. When a thug plays a nice guy, it initially falls in love, but when the nice guy replies with defection, the thug looks to players with higher expected payoffs. The thug will reach a point where all the other thugs have an expected payoff of 2.4, and the nice guys will have an expected payoff of 2.82, and thus the thug will next choose a nice guy whose expected payoff to the thug will rise to 2.874 and cause the thug to latch. The nice guy only chooses to play other nice guys, but can never refuse the latching thugs.

The interactions of this connected centers population can best be seen by looking at a specific example from an IPD/CR run. The two subpopulation sizes over 250 generations, and a significant play graph from a single generation are pictured in Figure 10.

Even though the nice guys would, by themselves, form a fully cooperating population, once the two behaviors are mixed together they form a long-term relationship from which the nice guys cannot easily escape. Note from Figure 10 that the more latchers on a nice guy, the higher its average score. This is because the extended cooperation balances the initial loss of points. A nice guy therefore depends on having thugs latch onto it in order to outscore the other nice guys. The more nice guys and fewer thugs, the better off a thug is, for the thug only reaps benefits from nice guys and loses points to other thugs. Thugs score higher as the number of nice guys in the population increases. Thugs thus start to displace nice guys from the population, but then their success drops and the number of nice guys eventually starts to increase. This creates a cycling in the composition of the population between nice guys and thugs, both of which depend on each other. This population behavior is similar to a predator prey relationship, except for the twist that, so long as nice guys exist, nice guys need to have thugs latch onto them in order to survive. Individuals who do not play one of the two strategies tend to do poorly, and invasion is generally prevented.

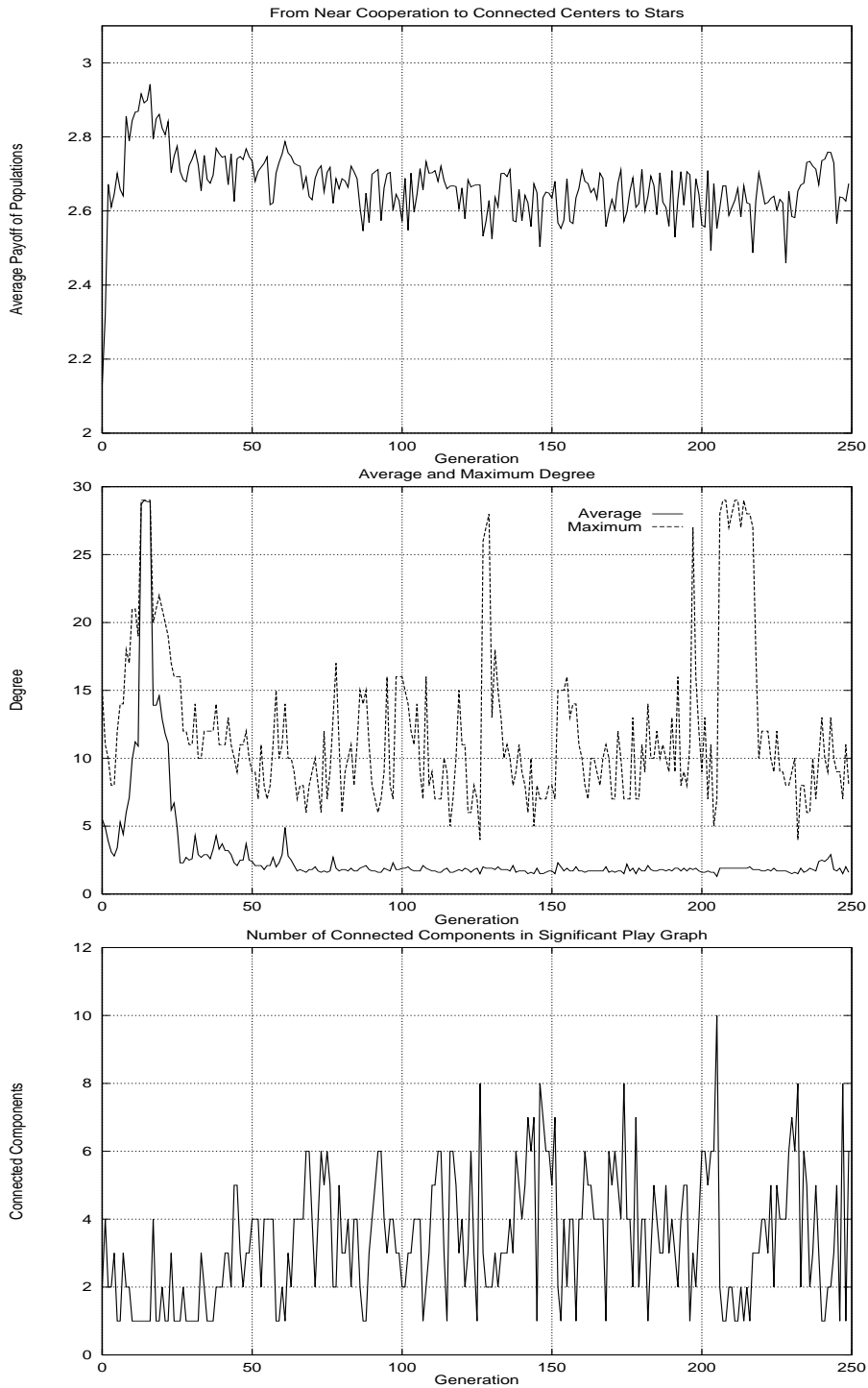

Figure 8: This collection of figures shows a run, minus mutants, which for all practical purposes reaches full nice cooperation early on and then goes into connected centers and then into stars behavior around generation 72. The connected centers population involves more defecting behavior than the example given in section 5.4.2. The run at times begins to exhibit more cooperative behavior, such as around generation 239, which is evidenced by the increase in the population's average payoff. The high peaks in maximum degree late in the run occured when there was temporarily only one hub and thus a single star.

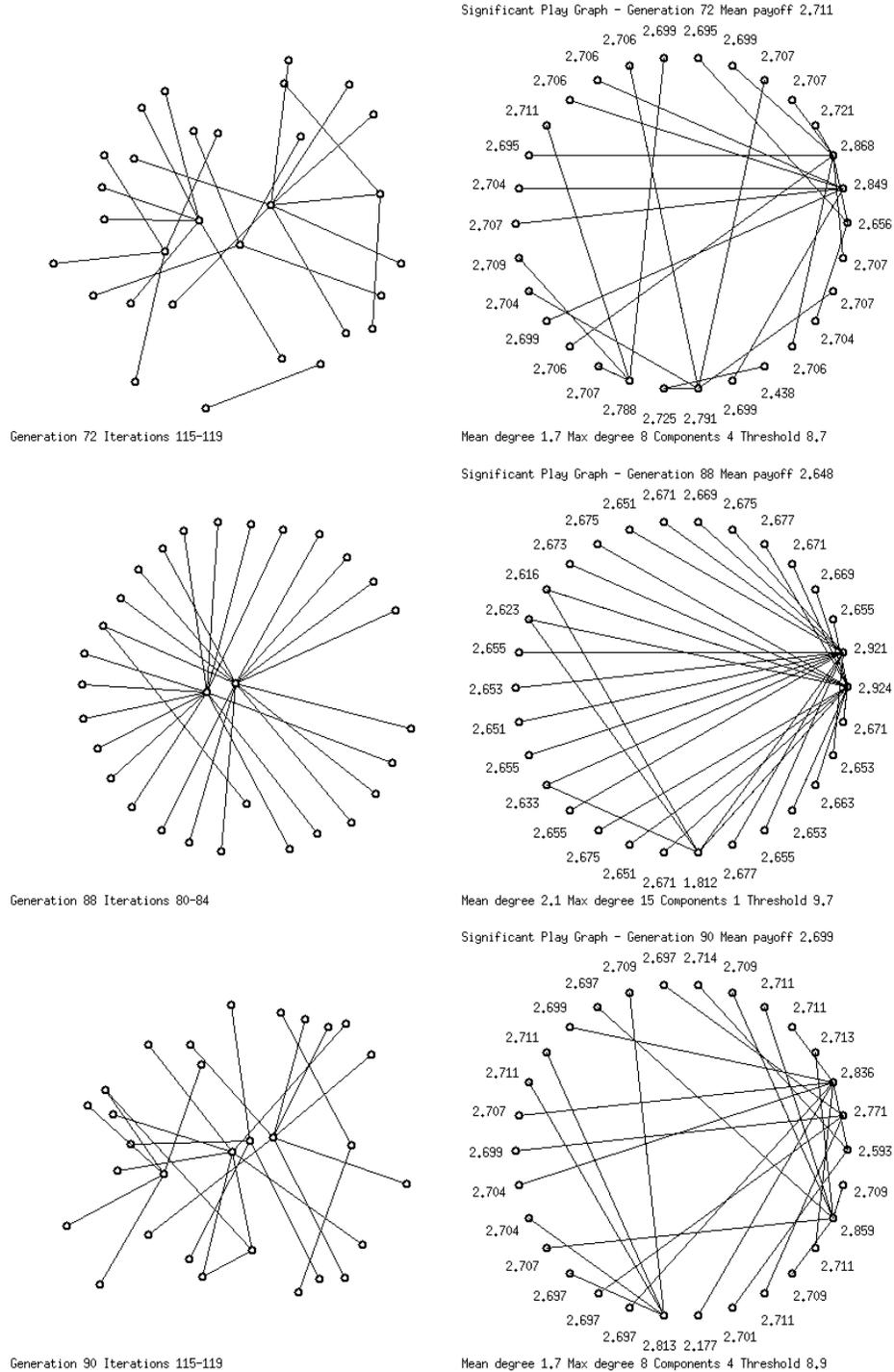

Figure 9: The stars nets. Notice how increasing number of stars "hub centers" only equalizes the fitness of the two types of individuals unlike Raquel and the Bobs where increasing numbers of Raquels cause Raquels' fitnesses to decrease at the expense of Bobs.

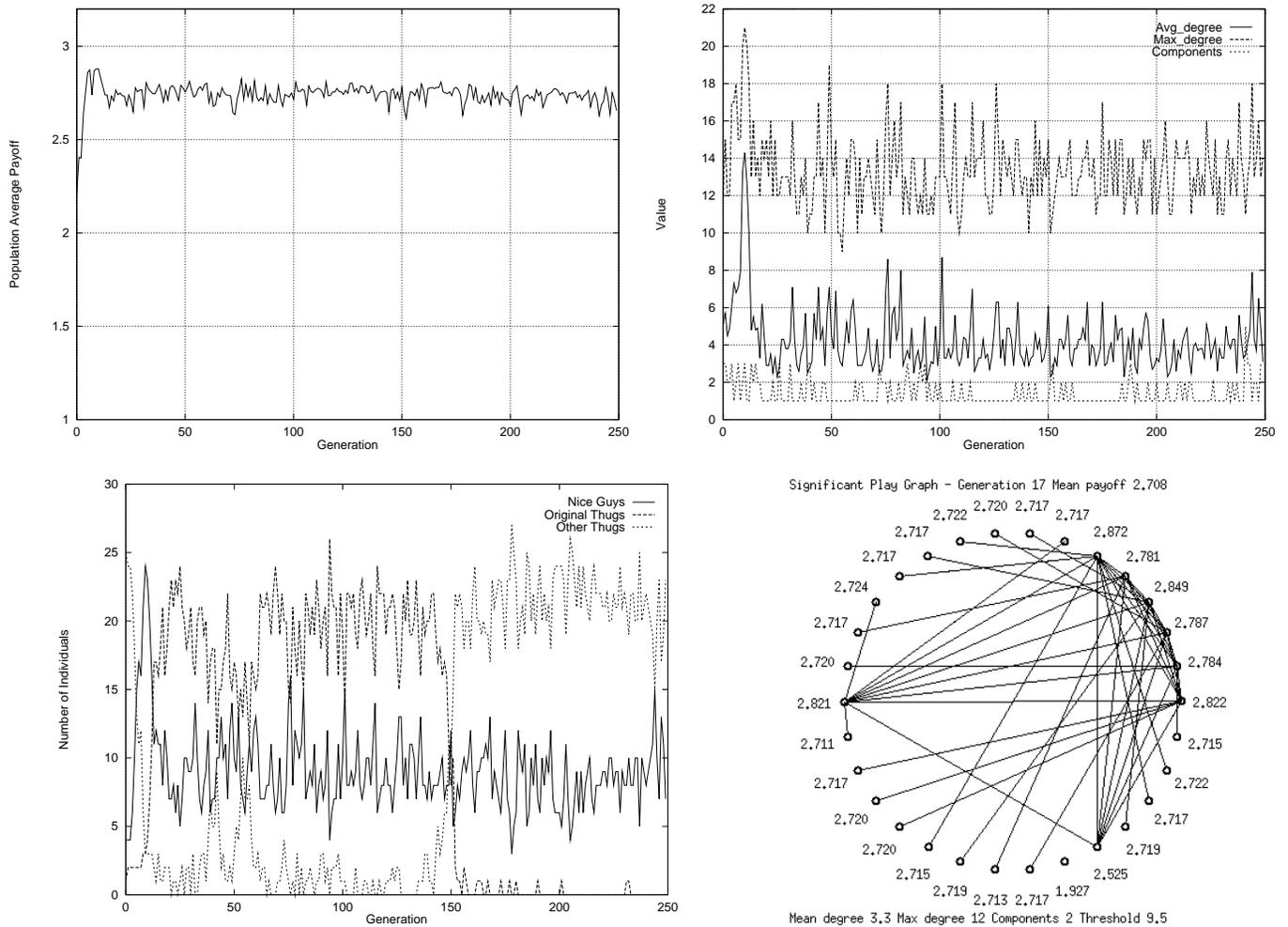

Figure 10: Plots from the sample connected centers population behavior described in the text. The upper right figure is of the population's significant play graph statistics. The lower left plot is of the number of nice guys, and thugs in the population v. generation, and the right figure is a sample significant play graph. The population of Nice Guys vs. Thugs mirror each other, i.e. as one goes up in size the other goes down in size. Note in the significant play graph that there is a group of eight players (nice guys) who all play each other and they each have varying numbers of thugs latched onto them. The cooperators which have more thugs latched on score higher.

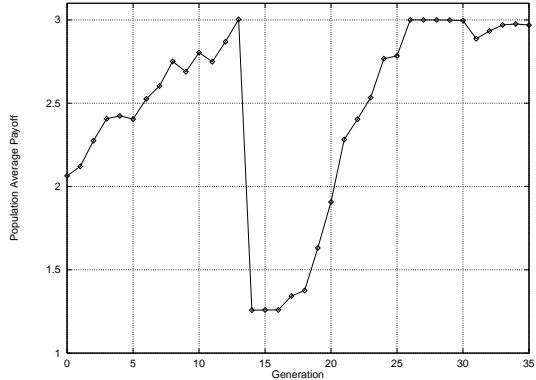

Figure 11: The population average payoff for the wallflower crash example given in section 6.1. The subsequent rise to cooperation is shown in Figure 12.

# 6 Crashes

Almost all of the IPD/CR populations run at the parameter settings of this paper, given in Table 1, evolve to have average fitnesses over 2.4 within a few generations, and then stay above this level. As we have noted earlier in the paper, populations often evolve which appear to be stable. However, evolution is always at work, and even fully cooperative populations can never be truly stable. While usually the transition out of a particular population behavior causes at worst a small shift downward in the population average of the average payoff, sometimes evolution causes a dramatic crash in this average payoff, and within one or two generations it drops more than half a point and stays down for several generations. These crashes appear to happen for reasons which are population specific, and the composition of the population may or may not be completely changed afterward. Below we describe an extinction and the resulting climb back out of a crash.

## 6.1 Wallflower

In the most dramatic of the crashes the population becomes a wallflower population. In a wallflower population every individual defects so much that everyone finds everyone else to be intolerable well before the end of the 150 iterations. Ideally, we would like the social network of this population to be a completely disconnected graph with no players connected to any other players. While it is true that everyone in most cases of a wallflower population plays everyone else equally (usually exactly four times at these parameter settings), the length of interaction is usually so short that it does not count as significant play relative to the number of iterations. The method we use to threshold the weighted play graph usually calculates the correct social network (completely disconnected) but sometimes when there is a single more cooperative mutant present it fails and produces a fully connected graph. The characteristic very low, flat, behavior of the average payoff in a wallflower population often precludes the need to look at the social network.

At these parameter values, wallflower populations do not last for many generations. Figure 12 shows the evolution of cooperation after a wallflower crash. The crash is caused by the extinction of five Raquel-like players at the end of generation 13. In generation 17, an individual is born that, after a few initial defections, can cooperate for the rest of the generation with one of the existing members of the population. In generation 18, the older of this pair has a child, that plays with the newer one in the same way as its parent. In generation 19, a player is born who is not only a nice cooperator with its parent, but who can also elicit cooperation with other members of the population. Both the latchers and the latchees do better than the rest of the population. In generation 20, a central group of three cooperators exists. The latchers are of two varieties. The first type alternates play between two cooperators, and the second type latches on, trades defections, and then cooperates. In generations 21–24, the more cooperative latchers beat out the other latchers, and the number of cooperative center players continues to grow. In generation 25, we see a large cluster of cooperators ostracizing some defectors. By generation 26 (not pictured), the entire population exhibits full nice cooperation.

## 7    Frequency of Population Behaviors

Our analyses of individual population behaviors in the previous two sections do not give any sense of how frequently each of the behaviors is observed. We have attempted to use our observations about the significant play graphs of our populations to develop automatic classification schemes, with very little success. As noted throughout our results section, there is significant overlap between the global statistics of each of the major population types. However, examining the distributions of the graph measures of the significant play graphs gives some sense of the frequency of different net structures. Figure 13 shows these distributions for the 100 simulations run for this paper. The results from the most common population behavior, full cooperation, are shown as dashed lines. Populations with a single component are the most common, even in populations which are not fully cooperative (both connected centers and Raquel and the Bobs populations tend to have a single component), large numbers of components are rare (even wallflower populations with 30 components are fairly uncommon), and there is a small peak centered at ten, due to the latching and stars populations. Notice that there is a gap in the average degree. Populations either have high or low degree. We can speculate that the peaks in the maximum degree are from latching at 3, from stars and connected centers at 13, and at 30 from R&B and full nice cooperation.

## 8    Non-Mutual Cooperation

Angeline [1] recently reported on a study in which he relaxed the $(L + H)/2 < C$ constraint on the payoffs in the Prisoner's Dilemma payoff matrix. With $(L+H)/2 > C$, the best payoff is still obtained by defecting against a cooperating opponent, and the dilemma still exists. However, now players who alternate defections and cooperations with each other achieve a higher average payoff than cooperation alone could produce. Now the truly cooperative action is to alternate defection and cooperation. It has been suggested that this is similar to a couple deciding that one would work while the other went to graduate school and then they would trade off again, with the end result being higher than if they had both simply continued in their present roles of simple cooperation. However, in order to alternate defections and cooperations, and be successful, players have to coordinate their actions over consecutive moves.

Figure 12: The evolution of cooperation after a wallflower crash. From right to left and from top to bottom: the generations described in section 6.1.

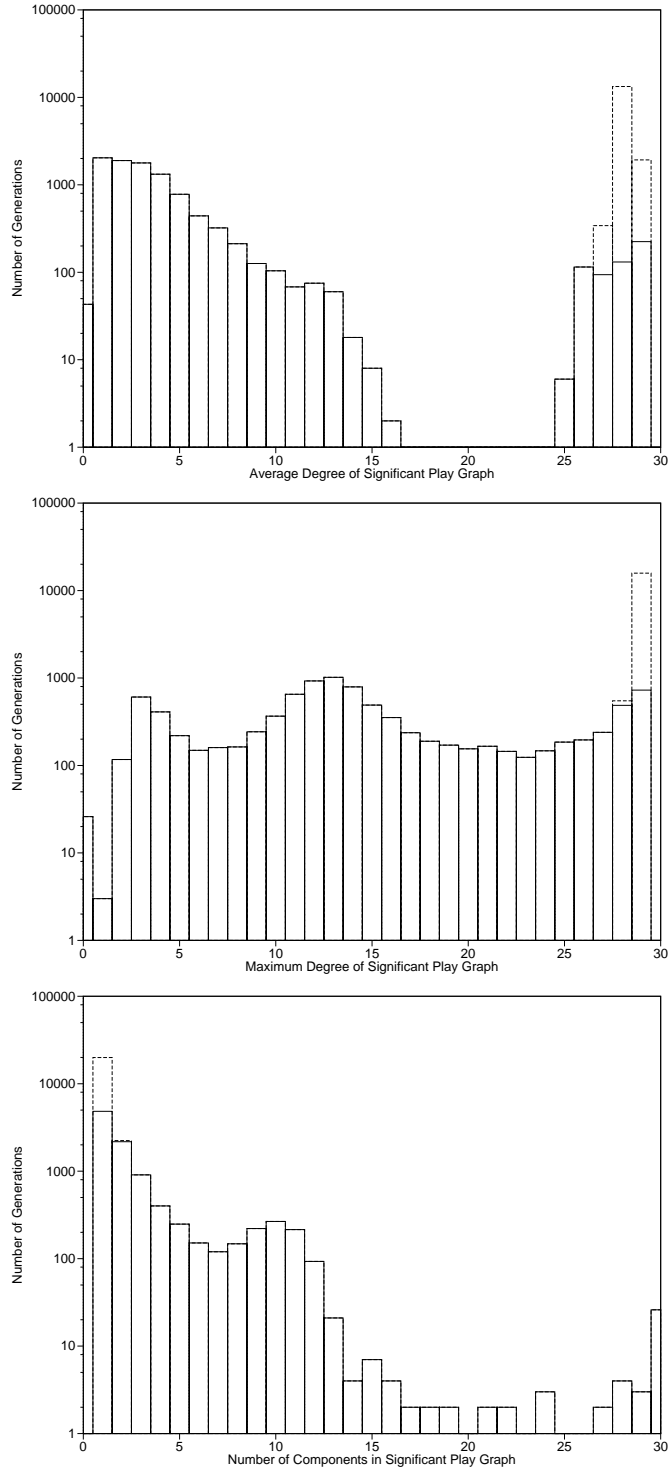

Figure 13: The distribution of graph metrics for 100 runs of 250 generations each. The dashed lines represent the full cooperation populations. A population was considered to be a full cooperation population if it had an average payoff greater than 2.88, average degree greater than 27, maximum degree greater than 27, and number of components less than or equal to three.

Angeline ran 10 runs of 200 generations with a population size of 100. Each player played a 100 iteration Iterated Prisoner's Dilemma with each of the other 99 players using the payoff matrix in Figure 1 with $H = 7$ instead of 5. A player's fitness was its average cumulative score over the 99 games. After an initial dip the average payoff across the 10 runs rose and at about the 75th generation settled down to an average around 3. Eight of his tens runs resulted in populations in which non-mutual cooperation dominated play.

Note that, for the types of finite state machines used by Angeline and in this paper, in order for non-mutual cooperation to arise it *requires* the coexistence of at least two different types of players in the population. It is not possible for a player to achieve non-mutual cooperation with a player identical to itself. But this means that in round robin Iterated Prisoner's Dilemma, as in Angeline, the players that are identical to each other have to find a way to cooperate with each other and even when they succeed in repeated mutual cooperation they must be content with the lower $C$ payoff. In our recreations of Angeline's simulations we found that the population often solves this problem by evolving into a number of subtypes which works to minimize the number of times individuals must use mutual in place of non-mutual cooperation. A subtype is considered to be a group of individuals with the same behavior. There exists a mix of subtypes with some having an initial move of cooperate ("nice" players) and others with defect as the initial move ("nasty"). When nice players play nasty players, they most often enter into immediate non-mutual cooperation of the type observed by Angeline (CDCD, DCDC). When two nice players play each other, they play the intial cooperate and then usually enter into a series of defections against each other, and if the players are different subtypes, then one will in a few moves play a cooperate against the other's defect. When this (CD,DC) event occurs, it triggers the players to enter into a non-mutual cooperation cycle. The different nice subtypes have varying lengths of defect moves after the intial cooperation. If both nice players play a cooperate after the series of defections, then they enter into mutual cooperation. The same process is used by nasty subtypes. Once again in an Iterated Prisoner's Dilemma simulation without choice and refusal, we see the use of the handshake/password mechanism mentioned in the introduction, but it relies on the long length of interaction between players.

When we did one hundred runs of IPD/CR at the standard parameters (Table 1) but set $H = 7$ as Angeline did, non-mutual cooperation immediately evolved in all but one run. In these runs, we saw two types of players which produced social nets similar to the one in Figure 14. We also saw a cycling of the numbers and amount of initial defectors and initial cooperators. In Figure 14 it can be easily seen that the population has two types of players present. The players with the lower scores around 3.1 are initial cooperators while the higher scoring players are initial defectors. Notice that the initial cooperators were the highest scorers in the *previous* generation as their position in the circle denotes. Two of the players' finite state machines can be seen in Figure 15. This cycling occurs because there is a conflict present when two strategies which play non-mutual cooperation coexist, since one strategy starts with a defection and the other starts with a cooperation. In IPD/CR the number of games any two pair plays is variable. If the length of interaction isn't an even number the initial defector has an advantage against the initial cooperator. Also, the initial defector possibly has an advantage against mutants. On the other hand, the initial defectors tend to take more mutual defection hits off of each other, especially as their population size increases.

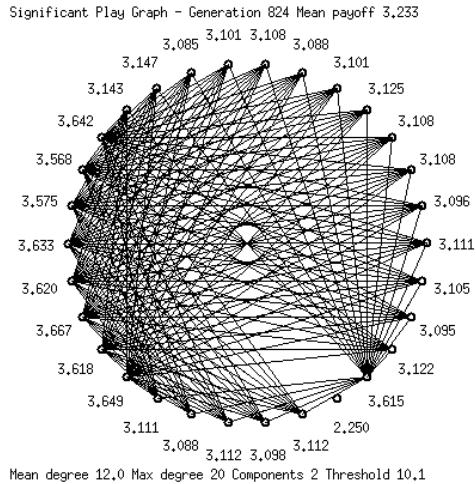

Figure 14: A social net showing non-mutual cooperation.

## 9  Future Ideas

The use of ideas from social networks and graph theory has given us a wealth of insight into the behavior of IPD/CR populations and the connection between individual actions and global societal properties. However, at the moment this analysis relies upon careful examination of the plots of the average fitness and significant play graph from each run, and, for noncooperative populations, visually examining the networks as the simulation runs. We would like to develop more automatic classification schemes which will give us statistics on the frequency of different population behaviors and alert us when novel phenomena are observed. This will be especially important for larger and more complex populations. Part of what is observable to the human eye is the time-dependence of the patterns in both the global statistics and the snapshots. Incorporating ideas about time-dependent graphs may help us with this latter problem. As well, the human eye can discern spatial patterns which are not obvious in the statistics. As we examine modifications to IPD/CR, we would like ideally to have tools which are not specific to the currently observed population behaviors which will allow us to automatically discern recurring patterns. We believe that tools of this nature are needed for artificial life multiagent systems in general, since one of the limits on using artificial life to model real-world biology is the difficulty of recognizing useful patterns.

In all studies of IPD/CR to date, we have used a single population in which all players have the potential to interact with all other agents. This seems reasonable when the population is small. However, in most populations individuals do not interact with all other individuals. Often there is some kind of spatial structure, and individuals interact only with those individuals they encounter. In territorial animal populations, these may be nearest neighbors, but in many populations interactions occur in groups, and individuals may move between groups. We plan to explore the impact of different spatial structures on the behavior

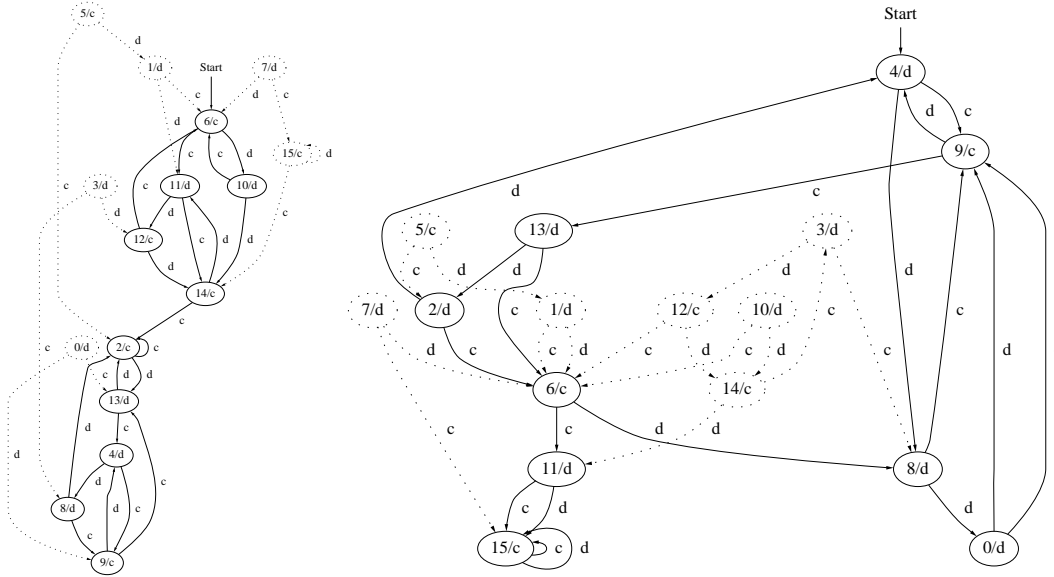

Figure 15: Two Moore machines from the non-mutual cooperative population shown in Figure 14. The dotted states represent states in the genetic code but they are unreachable by any input combination. The left figure shows the structure of the initial cooperator with a payoff of 3.147 in Figure 14. The right figure shows the structure of the initial defector with a payoff of 3.667 in Figure 14.

of IPD/CR.

The partner selection mechanism in IPD/CR is a strong determinant of the individual's behaviors. We have used expected payoffs, with exact ordering of potential partners, and a sharp tolerance cutoff. In Ashlock et al. [2] we presented preliminary results on the impact of allowing two of the parameters, $\tau$ and $\omega$, to evolve along with the finite state machines. These simulations indicated that many of the same social networks continue to emerge, and wallflower ecologies are very common. We plan to take this work further, allowing all of the choice parameters to evolve. However, it may be the hard cutoffs which allow such interesting behaviors as Raquel and the Bobs and Stars to emerge. Our individuals can distinguish between individuals with any tiny difference in expected payoffs, down to machine-precision. What happens if there is some fuzziness in their decisions, and individuals only use their expected payoffs to determine the probability of choosing or rejecting another individual? We will examine this issue in a number of different ways, for example by assuming that there is some noise involved in evaluating the expected payoffs.

IPD/CR was developed as a toy model which could be compared to existing results on prisoner's dilemma. It has yielded a rich variety of interesting results, but it is important to keep in mind that it does not model any particular system. In order to move a step closer to the real world, we plan to give individuals unique markers so that individuals can die and be born at random times, instead of in discrete generations, and preexisting individuals can retain their memory of each other. We also plan to give individuals visible markers, which evolve along with their strategies, which others can use to assess their "attractiveness" as partners even when they are strangers.

# 10 Conclusion

IPD/CR allows the formation of self-organizing social networks among selfish players. A wide variety of population structures evolve in our IPD/CR simulations, many of which have distinctive social network structures which arise from interesting interactions between individuals. The play graph is an effective tool for the quick diagnosis of these population behaviors when used in conjunction with other population statistics and visualization of the system. In this paper we have focussed on a subset of populations which have clear, distinct, social network structures, and which frequently occur for the particular IPD/CR environment of this paper. Table 6 summarizes their significant play graph statistics. These behaviors are not the only ones observed for the IPD/CR environment of Table 1, but they are prevalent in most of the IPD/CR environments we have studied in previous papers.

The significant play graph adds important information about the social behaviors of our system. However, the statistics we are currently collecting about these play graphs are not sufficient for us to automatically classify populations. This is partly a measure of the crudeness of these measures, and partly a reflection of the inadequacy of the concept of a static combinatorial graph. Not only do populations temporarily exhibit different behaviors because of mutants who survive for only brief periods of time, but even a single generation of IPD/CR play can be extremely complex with waves of rejection and momentary attacks on certain players. We have restricted our discussion in this paper primarily to populations which form their final social networks early in the generation, largely because populations which shift networks later in the generation are rare. However, they do arise, and as we develop more realistic models we will need to have ways of measuring and classifying more complex behaviors.

## Acknowledgments

Special thanks to Leigh Tesfatsion, who helped develop the IPD/CR model, and with whom we have had numerous thoughtful discussions concerning the social networks. Many thanks to the Iowa State Physics and Astronomy Department, especially the Gamma Ray Astronomy research group, for providing computer facilities and office space to Mark Smucker. This research was partially supported by an Iowa State University Research Grant funded under DHHS Grant # 2SO7RR07034-26 and by the Los Alamos Center for Nonlinear Studies.

| Hub | cdcd... | Hub | cdccc... | Spoke | dcdc... |
|-----|---------|-------|----------|-------|---------|
| Hub | cdcd... | Spoke | ddccc... | Spoke | dcdc... |

Table 4: The behaviors of the example disconnected stars run described in section 5.4.1.

| Nice guy | ccccc... | Nice guy | cdccc... | Thug | dcccc... |
|----------|----------|----------|----------|------|----------|
| Nice guy | ccccc... | Thug | ddccc... | Thug | dcccc... |

Table 5: From the example run discussed in the text, this table shows the behaviors of the two types of players in a connected centers population.

| Components | Average Degree | Maximum Degree | General Population Behavior |
|------------|----------------|----------------|----------------------------|
| 1 | 29 | 29 | Full Cooperation |
| 3–12 | 1–2 | 2–10 | Latching |
| 1 | 1–4 | 27–29 | Raquel & the Bobs |
| 1–8 | 1–3 | 5–9 | Disconnected stars |
| 1–2 | 3–5 | 10–15 | Connected Centers |
| 30 | 0 | 0 | Wallflower |

Table 6: Summary of general characteristics for the thresholded undirected play graph. Note that the numbers are only approximations.